\documentclass[letterpaper,journal]{IEEEtran}
\usepackage[T1]{fontenc}
\usepackage{amsmath,amsfonts}
\usepackage{graphicx,xcolor,tikz,url,cite}
\usepackage{balance}
\usetikzlibrary{arrows.meta,positioning,fit,calc}
\usepackage[hidelinks]{hyperref}
\definecolor{paperGreen}{HTML}{5F9500}
\definecolor{paperCharcoal}{HTML}{343434}
\definecolor{paperGray}{HTML}{777777}
\definecolor{paperLight}{HTML}{F2F4F2}
\definecolor{paperGrid}{HTML}{DFE4DF}
\tikzset{
 paperText/.style={font=\sffamily\fontsize{8.5}{10.5}\selectfont,text=paperCharcoal,align=center},
 paperHead/.style={font=\sffamily\fontsize{9}{11}\selectfont\bfseries,text=paperCharcoal,align=center},
 paperBox/.style={draw=paperCharcoal,line width=.6pt,fill=white,inner sep=4pt,align=center,
  font=\sffamily\fontsize{8.5}{10.5}\selectfont,text=paperCharcoal},
 paperArrow/.style={draw=paperCharcoal,line width=.75pt,-{Latex[length=1.9mm,width=1.2mm]}},
 paperGuide/.style={draw=paperGray,line width=.5pt,dashed}
}

\makeatletter
\renewcommand{\Huge}{\@setfontsize{\Huge}{21}{25pt}}
\makeatother
\begin{document}
\title{From Radio Measurements to Sensing Services: A 3GPP\\Perspective on Multi-UAV Tracking with Networked ISAC}
\author{Xingqin Lin}
\maketitle
\begin{abstract}
Integrated sensing and communication (ISAC) can turn radio measurements into networked sensing services, but accurate position estimates alone do not ensure useful tracking. Missed targets, late acquisition, track interruptions and aging observations also shape the information available to an application. This article examines tracking of noncooperative unmanned aerial vehicles (UAVs) from a Third Generation Partnership Project (3GPP) perspective. It connects sensing-service requirements and network architecture to three information interfaces: receiver search regions, measurement context and track validity. A five-UAV case study in a 3GPP urban scenario and a ray-traced San Francisco scene illustrates the tradeoff between target coverage and extra outputs, the distinction between acquisition and ongoing continuity, and the value of testing a static mission envelope before adding adaptation. An observation-age example explains how a state predicted to the present can still rely on old evidence. The resulting lessons are to align receiver search with requested coverage, preserve measurement time and uncertainty, and assess coverage, ambiguity and timeliness together. We conclude by pointing out fruitful avenues for further work in sensing design, calibration and service validation.
\end{abstract}

\section{Introduction}
\IEEEPARstart{P}{recise} position estimates can coexist with missing objects, interrupted tracks and outdated observations. A system may localize the objects it sees accurately while excluding others or retaining predictions after their supporting observations become too old. Before using a delivered coordinate, an application therefore needs to know what the receiver can examine, what the measurements support, and whether the estimate is still useful at the time of use. These questions connect sensing and localization to networking, robotics and edge processing.

Unmanned aerial vehicle (UAV) surveillance provides a concrete setting for examining them. A perimeter operator needs a current location, a consistent track and the age of its supporting observations. Integrated sensing and communication (ISAC) connects waveforms, resources and processing to communication and sensing objectives~\cite{liu2022}. Here, ground infrastructure senses external, noncooperative UAVs from radio reflections. Sites send measurement reports to a central tracker, which associates them across sites and time to estimate positions and motion. A track is that evolving estimate and its history under an internal label. The tracker infers how many targets are present and where they move.

Several parts of this problem are established. A tutorial on sensing in complex environments examines UAV detection and cooperative localization~\cite{lin2026wireless}. This article focuses on the subsequent temporal questions of track acquisition, continuity and observation age. Cooperative perceptive mobile networks address infrastructure UAV surveillance~\cite{zhang2024pmn}, and networked tracking and handover are explicit research topics~\cite{zhao2025handover}. An uplink sensing simulation testbed examines detection, association and tracking for a command-and-control consumer~\cite{gurung2026}. This article concerns the information that must survive between such components.

The estimation and control literature supplies essential foundations. Complementary radar views improve spatial information~\cite{haimovich2008}; multitarget filters represent uncertain existence and association~\cite{pmbm2018}; joint target--sensor-state methods address uncertain geometry~\cite{frohle2019}. Search-and-track sensor management already handles limited fields of view and demonstrates why small localization errors must be assessed alongside missed, false and lost targets~\cite{bostrom2021}. Complete target accounting and search-region management motivate the systems perspective developed here.

This article connects Third Generation Partnership Project (3GPP) sensing-service requirements and architecture to the decisions made by receivers, fusion processes and applications. It develops a systems perspective around three information interfaces: the receiver's search region, measurement time and uncertainty, and the validity of delivered states. A five-UAV case study in a 3GPP urban scenario and a San Francisco scene compares receiver search rules using the same received signals within each episode. The evaluation considers target coverage, extra outputs and complete scenes, while separating initial acquisition from subsequent track continuity. A separate static-region control examines whether an evolving search region is needed for the observed gains. Finally, an observation-age example connects track prediction to application use. Together, these contributions explain how receiver choices and the information preserved between components shape a sensing service, and motivate joint evaluation of coverage, uncertainty and timeliness.

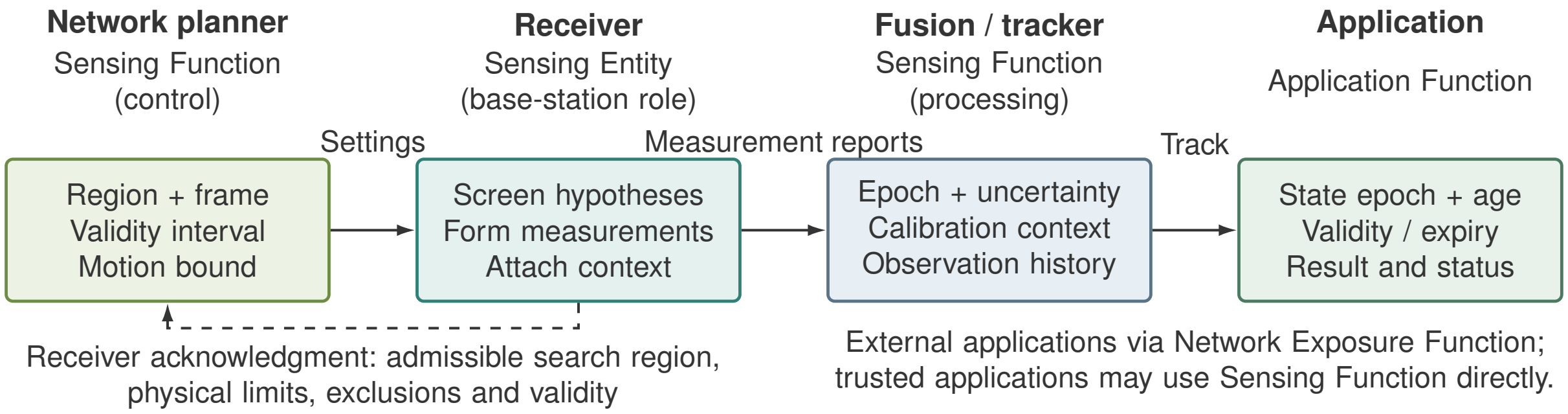
\begin{figure*}[t]
\centering
% Functional processing view with full sensing-role names; arrows show information flow.
% This is not a literal protocol topology or a specification of interface ports.
% Figure-specific flat palette; shared style and other figures are unchanged.
\begingroup
\definecolor{figThirtyOnePlanner}{HTML}{6D8D45}
\definecolor{figThirtyOnePlannerFill}{HTML}{ECF2E4}
\definecolor{figThirtyOneReceiver}{HTML}{2F8178}
\definecolor{figThirtyOneReceiverFill}{HTML}{E4F1EE}
\definecolor{figThirtyOneFusion}{HTML}{5A7487}
\definecolor{figThirtyOneFusionFill}{HTML}{E8EFF4}
\definecolor{figThirtyOneApplication}{HTML}{4C7A61}
\definecolor{figThirtyOneApplicationFill}{HTML}{E8F2EB}
\begin{tikzpicture}[x=1cm,y=1cm,
 figThirtyOneBox/.style={paperBox,line width=.9pt,rounded corners=2pt},
 figThirtyOnePlannerBox/.style={figThirtyOneBox,draw=figThirtyOnePlanner,fill=figThirtyOnePlannerFill},
 figThirtyOneReceiverBox/.style={figThirtyOneBox,draw=figThirtyOneReceiver,fill=figThirtyOneReceiverFill},
 figThirtyOneFusionBox/.style={figThirtyOneBox,draw=figThirtyOneFusion,fill=figThirtyOneFusionFill},
 figThirtyOneApplicationBox/.style={figThirtyOneBox,draw=figThirtyOneApplication,fill=figThirtyOneApplicationFill}]
\node[paperHead,inner sep=0pt] at (1.95,3.30) {Network planner};
\node[paperText,inner sep=0pt,text width=3.7cm] at (1.95,2.70)
 {Sensing Function\\(control)};
\node[paperHead,inner sep=0pt] at (6.15,3.30) {Receiver};
\node[paperText,inner sep=0pt,text width=3.7cm] at (6.15,2.70)
 {Sensing Entity\\(base-station role)};
\node[paperHead,inner sep=0pt] at (10.35,3.30) {Fusion / tracker};
\node[paperText,inner sep=0pt,text width=3.7cm] at (10.35,2.70)
 {Sensing Function\\(processing)};
\node[paperHead,inner sep=0pt] at (14.55,3.30) {Application};
\node[paperText,inner sep=0pt,text width=3.7cm] at (14.55,2.70)
 {Application Function};
\node[figThirtyOnePlannerBox,text width=3.0cm,minimum width=3.3cm,minimum height=1.45cm] (request) at (1.95,1.2)
 {Region + frame\\Validity interval\\Motion bound};
\node[figThirtyOneReceiverBox,text width=3.0cm,minimum width=3.3cm,minimum height=1.45cm] (rx) at (6.15,1.2)
 {Screen hypotheses\\Form measurements\\Attach context};
\node[figThirtyOneFusionBox,text width=3.0cm,minimum width=3.3cm,minimum height=1.45cm] (fusion) at (10.35,1.2)
 {Epoch + uncertainty\\Calibration context\\Observation history};
\node[figThirtyOneApplicationBox,text width=3.0cm,minimum width=3.3cm,minimum height=1.45cm] (app) at (14.55,1.2)
 {State epoch + age\\Validity / expiry\\Result and status};
\draw[paperArrow] (request.east)--(rx.west);
\draw[paperArrow] (rx.east)--(fusion.west);
\draw[paperArrow] (fusion.east)--(app.west);
\node[paperText,inner sep=0pt] at (4.05,2.08) {Settings};
\node[paperText,inner sep=0pt] at (8.25,2.08) {Measurement reports};
\node[paperText,inner sep=0pt] at (12.45,2.08) {Track};
% Geometry correction: keep the acknowledgment route above both text lines.
\draw[paperArrow,dashed] (rx.south)--(6.15,.20)--(1.95,.20)--(request.south);
\node[paperText,text width=7.8cm] at (4.05,-.30)
 {Receiver acknowledgment: admissible search region,\\physical limits, exclusions and validity};
\node[paperText,text width=7.7cm] at (12.45,-.15)
 {External applications via Network Exposure Function;\\trusted applications may use Sensing Function directly.};
\path[use as bounding box] (0,-.75) rectangle (16.6,3.55);
\end{tikzpicture}
\endgroup
\caption{Functional processing view mapped to sensing-service roles. The illustrated search-region, measurement-context and evidence-age fields are implementation choices.}
\label{fig:interfaces}
\end{figure*}

\section{From Service Requirements to Information Interfaces}
3GPP includes ISAC in Release 20~\cite{lin2025rel20}. Technical Specification (TS) 22.137 sets sensing-service requirements, including requested areas, continuity, contextual information, accuracy, latency and refreshing rate~\cite{ts22137}. TS 23.137 provides the Stage 2 architecture and service procedures. Its Release 20 scope is aerial-object detection and tracking using 3GPP base stations, referred to as gNBs~\cite{ts23137}. These specifications establish the service expectations and functional roles within which receiver, fusion and application choices operate.

In Fig.~\ref{fig:interfaces}, the application corresponds to the Application Function (AF), and the receiver to a gNB acting as a Sensing Entity. The Sensing Function (SenF) coordinates sensing and generates results; the planner and tracker illustrate its control and processing responsibilities. Multisite result generation is left to implementation. AFs outside the trusted domain access SenF through the Network Exposure Function; trusted-domain AFs may interact directly~\cite{ts23137}.

\subsection{Receiver Search Region}
The \emph{receiver search rule} determines which candidate ranges, directions and motions are examined or refined. A candidate is a possible interpretation of an echo; a measurement report is a retained estimate with its time and uncertainty. The search constraints define an \emph{admissible hypothesis set}, whose spatial extent is the configured \emph{search region}. A UAV may lie within the receiver's physical field of view yet outside this region. Expanding the region can retain another interpretation of the same measured signals; it does not extend the physical field of view or supply a new observation. Multiple sites can reduce geometric ambiguity or supply another view~\cite{haimovich2008}. A candidate excluded by one receiver may still be retained by another, but downstream association cannot recover evidence discarded by every view.

TS 23.137 requests specify detection or tracking, area, optional altitude limits, a time window, and one-time, periodic or event-triggered reporting. SenF selects sensing gNBs and derives their parameters from the request~\cite{ts23137}. Altitude limits reference sea level; translation into local coordinates requires an explicit height convention. Motion bounds and expanding regions remain implementation choices.

The requested region corresponds to the target sensing service area in TS 22.137, which also allows operator-controlled exposure of service availability and estimated quality for an area and time~\cite{ts22137}. A planner expresses this region in a common coordinate frame, with a validity interval and motion assumptions. Each receiver translates it into local range, direction and motion constraints. Search and sensor management already address choices about where and when to sense~\cite{bostrom2021}. A proposed receiver acknowledgment makes those choices visible to the planner by identifying the portion the site can examine and any exclusions. The planner can then revise the request or seek another view. Even within an examined region, weak or obstructed echoes can leave UAVs without reports.

\subsection{Measurement Time and Uncertainty}
TS 22.137 identifies capture time as context that can accompany sensing results~\cite{ts22137}. At the fusion interface, useful context includes the coordinate frame and units, acquisition and availability times, uncertainty, calibration information and an observation identifier. Fusion uses each site's position and orientation to relate local measurements to a common world frame. Acquisition time identifies when the scene was observed, while observation identifiers help recognize repeated evidence. A report may arrive after the tracker has advanced beyond that acquisition time. Out-of-sequence estimation incorporates the report at the appropriate past epoch and propagates the correction forward~\cite{barshalom2002}. The resulting state can describe the present while still relying on an earlier observation.

A site measurement report is an input to the tracker; a service-result report is an output from SenF and may describe several objects. TS 23.137 distinguishes service-report identifiers from object identifiers assigned within a sensing service and includes optional location, detection-time and movement information~\cite{ts23137}. An object identifier does not authenticate a UAV. Detection time alone does not specify the latest observation supporting a retained track; the evidence-age rule needs that observation's time.

Uncertainty also has direction. Range uncertainty spreads a candidate position along the viewing direction; angular uncertainty spreads it sideways. The same angular error produces a larger position error at greater range. A covariance describes the spread and relationships between these errors, allowing fusion to give more weight to the directions that each report constrains well. Complementary site views can reduce uncertainty that remains large from a single viewpoint. Calibration belongs in the same chain: a site's orientation error can shift several reports in the same direction. Combining those reports as independent measurements can reinforce the shared bias instead of averaging it away. Joint target--sensor-state estimation addresses this dependence~\cite{frohle2019}; receiver and fusion models should represent uncertainty in site position, orientation and timing alongside measurement noise.

Processing placement determines which information survives. Exchanging signal samples gives a fusion center access to returns that local detection might otherwise discard, at a transport and synchronization cost. Exchanging measurement reports reduces that load while leaving association to the fusion center. Exchanging tracks reduces it further but also passes along earlier association decisions and shared priors. The case study uses measurement-report exchange. Its transport and processing requirements must be considered alongside the information retained for fusion.

\subsection{State Validity}
An operator advisory near critical infrastructure may cue an independent verification sensor. An \emph{illustrative} application profile might require 10 m three-dimensional accuracy at use and an accepted-observation age no greater than 0.5 s. Prediction advances a track using its motion model and propagates uncertainty; an accepted measurement corrects the prediction. For this profile, a delivered track would identify its state epoch and uncertainty, last accepted observation time, updated or predicted status, and expiry. A failed quality or age criterion would prompt reacquisition or independent verification.

Missing coordinates can reflect different outcomes. An \emph{empty result} reports zero objects detected in the area and period sensed, a valid TS 23.137 result. \emph{Unavailable service} can arise when no sensing gNB can be selected: SenF rejects the request before acceptance or reports status afterward~\cite{ts23137}. A \emph{stale track} retains an estimate whose evidence has expired under our application rule. An application can continue monitoring after an empty result, seek alternative coverage when unavailable, and request reacquisition or verification for stale tracks.

The distinction is between the time represented by an estimate and the time of the observations behind it. The \emph{state epoch} is the time for which position, velocity and uncertainty are estimated. Prediction can advance this epoch to the present without adding a measurement. In this application profile, the \emph{supporting-observation epoch} is the acquisition time of the latest accepted observation supporting the estimate; it advances when more recent evidence is incorporated. The \emph{use time} is when the application acts. Evidence age is measured from the supporting observation's acquisition to use, regardless of when the report arrives or the state is predicted.

For example, a report acquired at 1.0 s and delivered at 1.4 s is already 0.4 s old on arrival. Suppose the tracker predicts the state forward for a query at 1.6 s. The state epoch is then 1.6 s, while the supporting-observation epoch remains 1.0 s: the evidence is 0.6 s old and exceeds the 0.5 s age limit. Measuring from arrival would give only 0.2 s and incorrectly permit use. A newly accepted observation acquired at 2.0 s and available at 2.1 s has age 0.1 s; use can resume if the quality criterion also passes. At the study's maximum speed of 50 m/s, a UAV can travel 25 m between the 0.5 s scans, so accuracy at acquisition can differ substantially from accuracy at use.

\section{Five-UAV Case Study}
\subsection{Scenario and Acquisition}
\begin{figure}[t]
\centering
% Native single-column setup: 8.65cm nominal width, with unscaled8--8.5pt labels.
% The reference triangle is500m per side. The initial sector has radius
% 500/sqrt(3)m, width120degrees, yaw30degrees. Aircraft and routes are illustrative.
% Existing style14.tex supplies the publication palette and arrow library.
\begingroup
\begin{tikzpicture}[x=1cm,y=1cm]
\tikzset{
 setupText/.style={paperText,font=\sffamily\fontsize{8.5}{10.5}\selectfont},
 setupSite/.style={circle,fill=paperCharcoal,draw=paperCharcoal,inner sep=2.1pt},
 setupReport/.style={draw=paperCharcoal!75,line width=.65pt,densely dashed,
  -{Latex[length=1.6mm,width=1mm]}},
 setupRadio/.style={draw=paperGreen!85!black,line width=.75pt,
  -{Latex[length=1.6mm,width=1mm]}},
 setupHeading/.style={draw=paperCharcoal,line width=.5pt,
  -{Latex[length=1.1mm,width=.8mm]}}
}
% Reference geometry at scale0.0094cm/m, not a scaled two-column drawing.
\coordinate (s0) at (1.00,1.65);
\coordinate (s1) at (5.70,1.65);
\coordinate (s2) at (3.35,5.720319398);
\path[fill=paperGreen!9]
 (s0)--++(-30:2.713546265)
 arc[start angle=-30,end angle=90,radius=2.713546265]--cycle;
\draw[draw=paperGreen,line width=.7pt,dashed]
 (s0)--++(-30:2.713546265)
 arc[start angle=-30,end angle=90,radius=2.713546265]--cycle;
\draw[draw=paperGray!65,line width=.55pt] (s0)--(s1)--(s2)--cycle;
\node[setupText,fill=white,inner sep=1pt] at (4.20,1.43) {500\,m};

\path (s0) ++(-15:1.65) coordinate (ua);
\path (s0) ++(10:2.15) coordinate (ub);
\path (s0) ++(35:2.00) coordinate (uc);
\path (s0) ++(60:2.20) coordinate (ud);
\path (s0) ++(80:1.50) coordinate (ue);
\def\setupAircraftGlyph#1#2{%
 \begin{scope}[shift={(#1)},rotate=#2]
 \path[fill=paperGreen,draw=paperCharcoal,line width=.3pt]
  (.11,0)--(-.085,.07)--(-.04,0)--(-.085,-.07)--cycle;
 \draw[setupHeading] (.15,0)--(.46,0);
 \end{scope}%
}
\setupAircraftGlyph{ua}{20}
\setupAircraftGlyph{ub}{55}
\setupAircraftGlyph{uc}{-10}
\setupAircraftGlyph{ud}{105}
\setupAircraftGlyph{ue}{-30}

\draw[setupRadio] (s0) .. controls (1.05,2.45) and (1.70,3.10) .. (uc);
\node[setupText,text=paperGreen!80!black,fill=white,inner sep=.5pt]
 at (1.34,2.52) {PRS};
\draw[paperArrow,line width=.6pt]
 (uc) .. controls (2.40,2.15) and (1.65,1.95) .. (s0);
\node[setupText,fill=white,inner sep=.5pt] at (1.97,2.15) {Echo};
\node[setupSite] at (s0) {};
\node[setupSite] at (s1) {};
\node[setupSite] at (s2) {};
\node[setupText,anchor=east] at (.83,1.80) {Site 0};
\node[setupText,anchor=north] at (5.70,1.43) {Site 1};
\node[setupText,anchor=east] at (3.12,5.72) {Site 2};

\path (s0) ++(50:2.713546265) coordinate (sectorCallout);
\draw[draw=paperGray,line width=.45pt] (sectorCallout)--(4.82,4.04);
\node[setupText,anchor=west,align=left] at (4.96,4.04)
 {Initial sector\\$120^\circ$};
\node[circle,draw=paperCharcoal,line width=.75pt,fill=paperLight,
 minimum size=.70cm,inner sep=0pt,font=\sffamily\fontsize{9}{10.5}\selectfont\bfseries,
 text=paperCharcoal] (tracker) at (7.45,2.95) {T};
\draw[setupReport] (s0) .. controls (2.30,-.10) and (6.60,.15) .. (tracker.south west);
\draw[setupReport] (s1) .. controls (6.30,1.65) and (6.70,2.40) .. (tracker.west);
\draw[setupReport] (s2) .. controls (6.40,5.85) and (7.45,5.00) .. (tracker.north);
\node[setupText,fill=white,inner sep=1pt] at (6.84,4.73) {Reports};
\node[setupText,anchor=north] at (7.45,2.47) {Tracker};

\path[use as bounding box] (0,.20) rectangle (8.65,6.15);
\end{tikzpicture}
\endgroup
\caption{Schematic of three-site monostatic sensing with positioning reference signals (PRS) and report fusion. The initial mission sector is shaded.}
\label{fig:setup}
\end{figure}
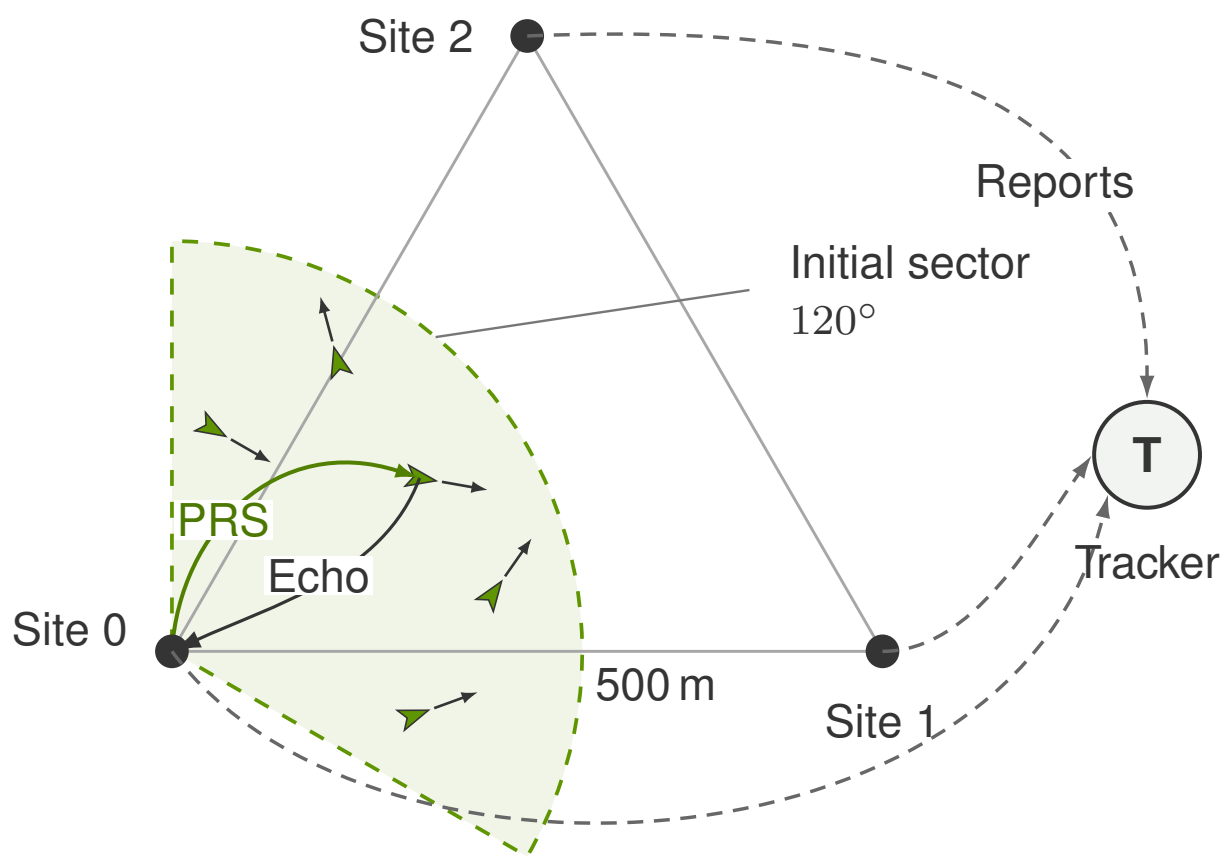

Each ground site has a co-located transmitter and receiver, a \emph{monostatic} view. Known positioning reference signals (PRS) illuminate the scene; reflected signals constrain target range, direction and radial motion. Each site converts retained echo candidates into time-stamped measurement reports. A central tracker combines the three sites' reports across scans to maintain three-dimensional position and velocity estimates. Figure~\ref{fig:setup} places this measurement-to-track chain in its physical geometry.

The study contains thirty episodes in each of two environments: a 3GPP urban scenario with macrocell sites and aerial targets, and an OpenStreetMap-derived San Francisco scene. Three sites form a 500 m horizontal triangle, with heights of 25 m in the 3GPP scenario and approximately 28--89 m in San Francisco. Five initial positions are drawn uniformly in area within a $120^\circ$ sector about site 0, with radius $R_0=500/\sqrt{3}\simeq289$ m and heights of 25--300 m. Horizontal speeds are uniform over 0--50 m/s, with uniform headings and constant height and velocity for six seconds. The flights maintain more than 10 m sampled target--target and target--site separation; San Francisco flights also avoid buildings. The results therefore describe short, separated, straight-line flights.

Thirteen scans start at 0, 0.5, \ldots, 6 s, with a common state epoch at each scan's acquisition center. In the 3GPP urban scenario, geometry and phase evolve while the initial line-of-sight condition, large-scale channel realization and path topology remain fixed~\cite{tr38901}. Propagation in the San Francisco scene is generated by ray tracing. These settings compare receiver policies under two propagation descriptions. Site positions, orientations and clocks are exact, sensing is isolated from communication interference, and report uncertainty is approximated.

Each site uses an $8\times8$ dual-polarized panel, about 32 MHz of disjoint PRS bandwidth at 4 GHz, and 47.2 dBm conducted transmit power. Sparse PRS transmissions are distributed within a nominal 32 ms acquisition window, with target-free observations for calibration. Refinement can localize an isolated return more precisely than the coarse search grid. Resolving nearby UAVs is a separate task: a 1 m localization tolerance does not imply 1 m target separation.

\subsection{Receiver and Tracker}
For a monostatic view, the receiver interprets echo delay as two-way propagation time and converts it to candidate range. Array response supplies a direction estimate, while Doppler supplies a radial-motion estimate along the site's line of sight. Range and direction locate a candidate relative to that site; radial velocity gives only one projection of its velocity. Combining views and successive scans provides information for estimating three-dimensional motion. Range/Doppler Fourier processing and consistency across antennas, transmission times and signal codes produce echo candidates. The same range/Doppler peak can correspond to several possible directions. The receiver search rule determines which directions proceed from the coarse grid to local range-and-angle refinement, producing different reports from the same signals.

Each measurement report contains range, azimuth (horizontal bearing), elevation (vertical angle) and radial velocity, with an approximate covariance based on signal-to-noise ratio and resolution. A central extended Kalman filter predicts three-dimensional position and velocity using a constant-velocity model. An associated report updates an existing track; an unassigned report may initiate a tentative one. Internal labels connect the estimates across time. A tentative track becomes \emph{confirmed} after three distinct-scan hits in four scans; maintenance needs two in three. Several site updates in the same scan still count as one hit. Confirmation measures report persistence; geometric agreement with a UAV is assessed separately during evaluation. A \emph{coast} is a predicted track state without a new accepted report in that scan. Position estimation, measurement updates and track confirmation are therefore separate stages.

\subsection{Receiver Search Policies}
The policies use the same received and calibration signals, detector, fitting methods and tracker. Transmit power, bandwidth and site placement are fixed. Only the search rule changes which candidate positions are considered and how local fitting is constrained.

\begin{itemize}
\item \textbf{Original guards:} a fixed disc of radius $R_0+5$ m about the initial sector origin, with site-dependent sector and height tests.
\item \textbf{Dynamic search region:} the initial sector expanded horizontally by $50t+5$ m, where $t$ is acquisition time since mission start, with a common world-height interval of 20--305 m. This region follows the mission's declared motion bound.
\item \textbf{Static search region:} the same harmonized sector/height rule, using the bound at the last acquisition time for the entire mission.
\end{itemize}

All policies retain the same range/angular processing limits and computation budgets. The sixty-episode comparison changes region shape and height consistency as well as time dependence, testing their combined effect on receiver search. Six scenarios from the main study, three per environment, are compared under all three policies. The static region admits the final-time envelope from the outset, while the dynamic region expands toward it. This control tests whether time dependence is needed for the observed benefit and is reported separately from the thirty-episode statistics per environment.

\begin{figure}[t]
\centering
\includegraphics[width=\linewidth]{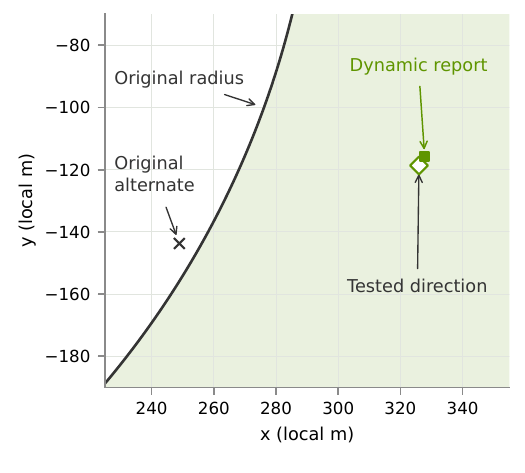}
\caption{Search-region example. The dynamic region retains a direction excluded by the original guard; shading marks the additional admissible region. The scan gains one match and one extra output.}
\label{fig:case}
\end{figure}

Figure~\ref{fig:case} illustrates how these rules affect a scan selected for its coverage gain. The original radius rule excludes a candidate direction while retaining another coarse interpretation. The dynamic region admits the excluded direction for refinement, allowing the resulting report to update an existing confirmed track. The scan gains one geometric match and one extra output: a broader search can recover useful evidence while also retaining competing interpretations.

\subsection{Evaluation Metrics}
The tracker associates measurement reports using its observations and predicted states. Evaluation then compares its confirmed tracks with the simulated UAV positions at each acquisition epoch. Each UAV contributes one required position and is \emph{matched} if a confirmed track lies within the stated three-dimensional tolerance. Matching is one-to-one: each track and each UAV can participate in at most one match. \emph{Matched coverage} is the fraction of required positions matched. Remaining confirmed tracks are \emph{extra outputs}, which can include spurious tracks or badly localized estimates of real UAVs. An accepted measurement update can still leave a track outside the evaluation tolerance.

For intuition, four matched UAVs out of five give 80 percent coverage but an incomplete scene. Five correct matches plus one extra output give 100 percent coverage and still fail the \emph{complete-scan} criterion: all five matched with zero extras. One unmatched track persisting for eleven scans contributes eleven extra outputs, not eleven distinct UAVs. Joint accounting for localization, missed targets and false targets is central to multitarget performance evaluation~\cite{gospa2017}; reporting coverage, extras and complete scans separately makes that tradeoff visible.

The first two scans precede three-hit confirmation. The remaining eleven scans form a fixed \emph{post-startup window}, approximately 1--6 s, including UAVs still awaiting acquisition. Each environment contributes 1,650 required positions and 330 scans to this window. All five UAVs remain in the denominator, including those never acquired. For an individual UAV, \emph{acquisition} means its first confirmed geometric match. Its subsequent history records later matched epochs, sampled gaps, label changes and coasts. A fresh output has an accepted report update in that scan. Never-acquired UAVs remain separate; those first matched at the final scan have no follow-up. Because the records are sampled every 0.5 s and end at 6 s, they describe sampled continuity over that observation period.

Service accuracy combines an error tolerance with a confidence level. The prespecified 1 m matching tolerance is the primary experimental criterion; a 5 m sensitivity view evaluates the same estimates with a looser position tolerance. Extra outputs count unmatched tracks, whereas the specification's false-alarm probability concerns binary decisions when a target is absent~\cite{ts22137}. The thirty episodes per environment are the statistical units for the 95 percent pointwise intervals, calculated using 2,000 whole-episode bootstrap resamples with policy results paired within each episode. This preserves the dependence between successive scans and the policies' shared flight and signal conditions. The six-scenario control is summarized separately.

\section{Tracking Performance}
\subsection{Coverage and Search Regions}
\begin{figure}[t]
\centering
\includegraphics[width=\linewidth]{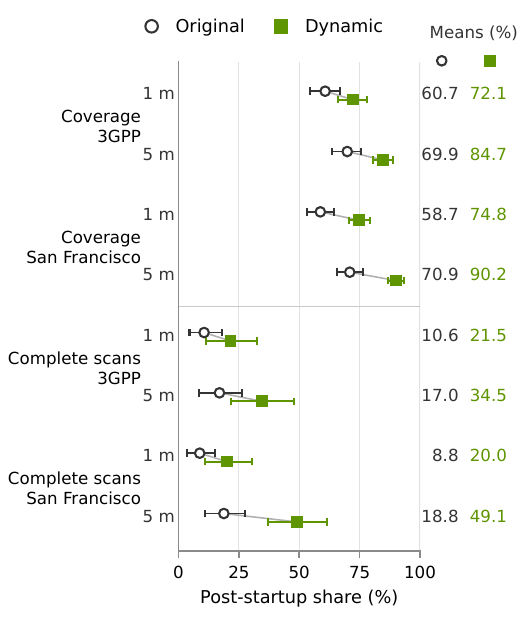}
\caption{Post-startup matched coverage and complete five-target/zero-extra scans at 1 m and 5 m tolerances. Whiskers show 95 percent whole-episode bootstrap intervals; each environment has 30 episodes, 1,650 required positions and 330 scans.}
\label{fig:tradeoff}
\end{figure}

Figure~\ref{fig:tradeoff} compares two views of performance: how often individual UAVs are matched, and how often the entire five-UAV scene is correct with no extra outputs. At the primary 1 m tolerance, the dynamic search region raises post-startup matched coverage from about 61 to 72 percent in the 3GPP scenario and from 59 to 75 percent in San Francisco. At 5 m, it reaches about 85 and 90 percent, respectively. These are the same estimated tracks assessed with a looser position criterion; the difference highlights estimates that are close to a UAV but fail the stricter criterion. The improvement between policies comes from changing which interpretations of the same signals reach the tracker. Receiver search design therefore belongs alongside waveform design, localization and association in the sensing chain.

Complete scenes remain much less frequent than individual matches. At 1 m, only about one in five post-startup scans contains all five targets with no extras under the dynamic policy. Even at 5 m, the corresponding shares are only about one third in the 3GPP scenario and one half in San Francisco. The broader search also increases the aggregate number of unmatched confirmed outputs, although the interval for the San Francisco change includes zero. Thus higher target coverage does not automatically yield an unambiguous scene: recovered matches and competing outputs must be assessed together.

\begin{figure}[t]
\centering
\includegraphics[width=\linewidth]{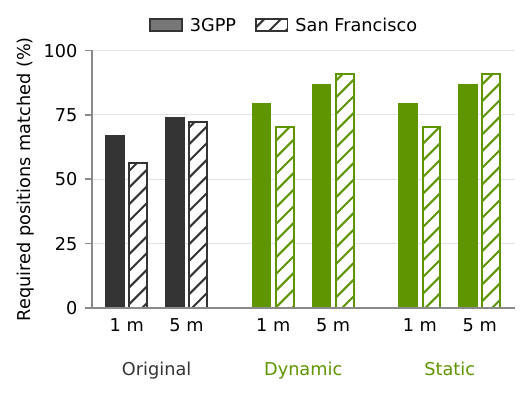}
\caption{Post-startup coverage at 1 m and 5 m tolerances in six scenarios, three per environment. Bars show means.}
\label{fig:comparison}
\end{figure}

Whether the search region needs to evolve is a separate design question. In the six-scenario control (Fig.~\ref{fig:comparison}), static and dynamic regions attain the same reported coverage, extra-output and complete-scan measures at every evaluated tolerance. An evolving boundary is therefore not needed to obtain the gains in these six cases. Time dependence would be useful if it improved the coverage--ambiguity tradeoff or reduced search effort.

\subsection{Acquisition and Continuity}
Continuity for moving-object sensing is a service expectation in TS 22.137~\cite{ts22137}. Aggregate coverage combines two different experiences: waiting for a target's first confirmed match and keeping it matched thereafter. A target acquired late but then tracked consistently can have the same overall coverage as one acquired early and repeatedly lost. The distinction matters to an operator deciding when surveillance becomes available and how dependable it remains.

The confirmation rule makes one second the earliest possible acquisition. With the dynamic region, roughly two thirds of the UAVs acquire a confirmed 1 m match at that first opportunity. Others take longer, and some have never attained a confirmed 1 m match by the six-second endpoint: 16 of 150 in the 3GPP scenario and ten in San Francisco. These targets are part of the all-target coverage in Fig.~\ref{fig:tradeoff}, even though they have no post-acquisition history to summarize.

\begin{figure}[t]
\centering
\includegraphics[width=\linewidth]{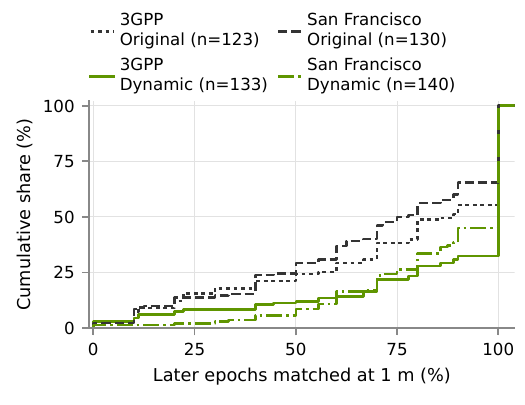}
\caption{Distribution of the percentage of subsequent scans with a confirmed 1 m match, for UAVs acquired before the final scan. Rightward curves indicate better continuity; $n$ is the number of UAVs with follow-up.}
\label{fig:continuity}
\end{figure}

Figure~\ref{fig:continuity} asks: \emph{once a UAV is acquired, how often is it matched at later scans?} Each acquired UAV contributes the percentage of subsequent scans with a confirmed 1 m match. A UAV matched in eight of ten later scans, for example, contributes 80 percent. At any horizontal-axis value, the curve gives the share of acquired UAVs with continuity at or below that value. A lower curve means fewer UAVs with continuity that poor; a rightward shift indicates improvement. The step at 100 percent represents UAVs matched at every subsequent scan.

The dynamic-policy curves show better continuity in both environments. The average later matched fraction is about 85 percent, compared with about 74 percent under the original guards in the 3GPP scenario and 70 percent in San Francisco. Each history begins at that UAV's acquisition, so the curves include different amounts of follow-up and somewhat different target populations. Never-acquired UAVs and those acquired only at the final scan are counted separately. Figures~\ref{fig:tradeoff} and~\ref{fig:continuity} together distinguish acquiring targets from retaining them.

Track labels remain stable between adjacent matched scans, but a few change across gaps. A gap can reflect a localization error or loss of confirmation even when the tracker retains a prediction. Most dynamic-policy matches receive a fresh report in that scan; coasting contributes little of the matched coverage. These observations point to improving measurement availability and interpretation alongside track-history management, rather than treating every interruption as a lost internal label.

The illustrative advisory profile connects tracking to application use. With its 10 m tolerance and 0.5 s observation-age limit, dynamic-policy coverage at acquisition epochs is about 90 percent in both environments. The age condition excludes no further matches here because confirmed coasts span at most one missed scan. Processing, transport and waiting can still consume the age budget before use. A state predicted to the delivery time can therefore rely on evidence that has become too old for the application, even if it met the criterion at acquisition.

\section{Network Design Implications}
\subsection{Coverage and Search Design}
The planner and receiving sites need to agree which part of a requested region each site can examine under its range, angle and processing limits. Declared exclusions let the planner revise the request or seek complementary views. This agreement creates an opportunity to detect targets; propagation, sensitivity and target reflections still determine which observations are obtained.

The same-signal comparison places search-region design alongside waveform and site selection. A useful interpretation of a received signal can be excluded by the search rule, while a broader region can retain competing interpretations as well as useful reports. Coverage, extra outputs and complete scenes must therefore be assessed together. Average error among matched tracks overlooks missing targets, while matched coverage alone overlooks competing outputs.

For a bounded mission, a compatible static search region provides a practical baseline. An evolving region should earn its added complexity through better coverage, fewer competing outputs or lower processing demand. That benefit depends on decisions such as which hypotheses are refined and how search effort is distributed across sites and time. The six-scenario result motivates testing this simpler baseline before adopting a changing region.

\subsection{Measurement Context and Timeliness}
The exchange level should preserve the information fusion needs within the delivery budget. Measurement reports leave association decisions to fusion; tracks also carry earlier decisions and shared priors. Repeated observations or shared site biases must be recognized to avoid overstating confidence. The exact site positions, orientations and clocks assumed in this study simplify that task; practical fusion also needs to account for uncertainty in those quantities. Preserving more samples offers little benefit if they cannot be processed before the application's deadline, while compressing observations into tracks can remove context needed to handle shared evidence. The exchange level, estimation method, transport capacity and delivery deadline should therefore be designed together.

TS 22.137 distinguishes service latency, measured from a triggering event to result availability at the sensing-system interface, from refreshing rate, which describes how frequently results are generated~\cite{ts22137}. The advisory application additionally tracks evidence age at use. A regular acquisition schedule does not guarantee regular result delivery: if processing takes longer than the interval between scans, a serial pipeline accumulates a queue. Parallel processing can improve throughput while individual results still arrive too late.

Sustaining two scans per second without queue growth and meeting an acquisition-to-use age budget are separate goals. The age budget includes receiver processing, transport, fusion and waiting before use. Further refinement is worthwhile when its estimation benefit justifies the delay. A prediction can be current while its observations are old, so the service also needs an explicit response to expiry: reacquisition, independent verification or a stale-state indication. Throughput, queue delay and evidence age together reveal whether processing meets the intended service needs.

\balance
\section{Conclusion}
This article connected 3GPP sensing-service expectations with receiver search, measurement context and state validity. The multi-UAV study illustrates why localization precision must be considered alongside missed UAVs, extra outputs and interruptions. Its six-scenario control motivates a mission-compatible static baseline before introducing adaptation. Acquisition histories and observation age add two further questions: when a useful track first becomes available, and how long its evidence remains useful. We conclude by pointing out some fruitful avenues that require further work.

\textbf{Search and resource design.} Jointly choosing site placement, sensing resources and hypothesis refinement could improve the balance between target coverage, competing outputs and processing cost. Adaptive search policies should be compared with a mission-compatible static baseline, with their additional complexity justified by better sensing performance or lower resource demand.

\textbf{Calibration and network fusion.} Practical fusion needs uncertainty models that account for shared calibration errors and dependence between sites, alongside each report's localization uncertainty. Online calibration of site positions, orientations and clocks, association under uncertain geometry, and recognition of shared observations are useful directions for improving the information passed to the tracker.

\textbf{Timely delivery.} Acquisition timestamps should be preserved through receiver processing, network transport and fusion so that throughput and evidence age can be assessed together. Joint sensing and communication scheduling could respond to expiring tracks, while applications distinguish an empty result, unavailable service and an outdated estimate.

\textbf{Service-level validation.} Longer observations of maneuvering UAVs, occlusion, clutter and changing propagation would test how well the design lessons extend to more varied conditions. End-to-end evaluation should also consider realistic traffic and processing loads, assessing whether delivered tracks meet the application's coverage, uncertainty and timeliness needs when a decision is made.

\end{document}